\documentclass[aps,prb,reprint,groupedaddress]{revtex4-1}

\usepackage{amsmath,graphicx,color}
\usepackage{bm}

\newcommand{{\todo}}[1]{{\color{red} \bf #1}}
\newcommand{\pd}{{\phantom{\dagger}}}
\newcommand{\up}{\uparrow}
\newcommand{\dw}{\downarrow}
\newcommand{\ket}[1]{\left|#1\right\rangle}
\newcommand{\bra}[1]{\left\langle#1\right|}

\begin{document}

\title{Dimensional Tuning of Majorana Fermions and the Real Space Counting of the Chern Number}

\author{Eric Mascot,$^{1}$ Sagen Cocklin,$^{1}$ Stephan Rachel,$^{2}$ Dirk K. Morr$^{1}$}
\affiliation{$^{1}$University of Illinois at Chicago, Chicago, IL 60607, USA}

\affiliation{$^{2}$School of Physics, University of Melbourne, Parkville, VIC 3010, Australia}

\date{\today}


\begin{abstract}
We demonstrate that it is possible to use atomic manipulation techniques to adiabatically tune between one-dimensional and two-dimensional topological superconductors in  magnet--superconductor hybrid (MSH) structures. This allows one to change the nature of the associated Majorana fermions between localized bound states and chiral Majorana edge modes, and provides a new approach for counting the topological invariant of the system -- the Chern number -- in real space. Moreover, we show that the topological nature of finite-size MSH structures can be characterized using the Chern number density which adiabatically connects MSH structures to macroscopic, translationally invariant topological superconductors characterized by an integer Chern number. These results open new possibilities for studying the nature of topological superconductivity in MSH structures using atomic manipulation techniques.

\end{abstract}

\maketitle

%
%
\section{Introduction}

The recent observations of Majorana modes in one-\cite{mourik-12s1003,nadj-perge-14s602,ruby-15prl197204,pawlak-15npjqi16035,kim-18sa5251} and two-dimensional\,\cite{menard16-nc2040,palacio-morales2019} topological superconductors hold the promise for topology-based technologies and topological quantum computation\,\cite{nayak-08rmp1083}. The realization of these technologies will not only require the ability to engineer Majorana fermions in nanoscale systems, but also to manipulate them spatially at the length scale of a few lattice constants. Magnet--superconductor hybrid (MSH) systems consisting of magnetic adatoms deposited on the surface of conventional $s$-wave superconductors represent promising candidate systems to achieve these goals. Indeed, single-atom manipulation techniques were recently employed to engineer topological superconductivity in MSH structures, where one-dimensional (1D) {\it Shiba} chains of magnetic Fe adatoms were built atom-by-atom on a superconducting Re surface using the tip of a scanning tunneling microscope\,\cite{kim-18sa5251}, allowing to visualize the emergence of Majorana bound states. Similarly, interface engineering has proven crucial in the creation of two-dimensional (2D) topological superconductivity and the direct visualization of chiral Majorana edge modes\,\cite{palacio-morales2019} in MSH structures consisting of {\it Shiba} islands of magnetic adatoms on the surface of $s$-wave superconductors. It is this ability to engineer real space structures which likely holds the key to the realization of topological quantum devices.

The possibility of using atomic manipulation and interface engineering techniques to realize  Majorana fermions has raised a series of intriguing questions, whose answers have the potential to provide unprecedented, fundamental insight into the nature of topological superconductivity. First, is it possible to adiabatically tune between topological phases in 1D and 2D MSH structures without undergoing a phase transition? The answer to this question is of particular interest not only because topological superconductors in 1D and 2D are in different homotopy groups -- with homotopy group $Z_2$ in 1D, and $Z$ in 2D\,\cite{schnyder-08prb195125,kitaev09}  -- but also because it would potentially open new possibilities to tune the nature of Majorana fermions between localized bound states and delocalized chiral edge modes. Second, can one identify the topological invariant of 2D topological superconductors -- the Chern number $C$ -- in real space? The fact that $C$ is encoded in real space at the edges of 2D Shiba islands as the number of chiral Majorana edge modes suggests that the real-space identification of $C$ might indeed be possible.

In this article, we will address these questions and demonstrate the ability -- using atomic manipulation techniques -- for adiabatic tuning between chiral Majorana edge modes and localized Majorana bound states in MSH structures.  Specifically, by attaching networks of Shiba chains to Shiba islands, one can arbitrarily move Majorana fermions between the edge of the island, and junctions or end points in the chain networks. Moreover, we show that it is this ability that opens a new approach to identifying the Chern number in real space by counting the number of Majorana bound states in the attached networks.  Finally, we demonstrate that the topological nature of finite-size MSH structures can be characterized using the Chern number density which adiabatically connects MSH structures to macroscopic, translationally invariant topological superconductors characterized by an integer Chern number. This demonstrates that even finite size 2D MSH systems, such as the ones realized experimentally \cite{menard16-nc2040,palacio-morales2019}, can be in a well-defined topological phase, exhibiting $|C|$ chiral edge modes. Our results open new possibilities for the study of topological superconductivity in MSH structures using current atomic manipulation techniques and open new avenues for the creation of the first Majorana-based quantum devices.

%
%
\section{Theoretical Model}
To investigate the engineering of Majorana fermions, we consider MSH structures in which magnetic adatoms are placed on the surface of a conventional $s$-wave superconductor with a Rashba spin-orbit interaction. Such systems are described by the Hamiltonian
\begin{align}
H &= \sum_{{\bf r},{\bf r^\prime}, \sigma} \left( -t - \mu \delta_{\bf r,r^\prime} \right) c_{{\bf r} \sigma}^\dag c^\pd_{{\bf r^\prime} \sigma} + \Delta \sum_{\bf r} c_{{\bf r} \up}^\dag c^\dag_{{\bf r} \dw} + {\rm H.c.} \nonumber \\
& +\,i \alpha  \sum_{{\bf r}, {\bm \delta} ,\alpha, \beta} c_{{\bf r},\alpha}^\dag \left( {\boldsymbol{\delta}}
\times {\boldsymbol \sigma}\right)^z_{\alpha \beta} c^\pd_{{\bf r}+\boldsymbol{\delta},\beta} \nonumber \\
 & + J  \sum_{{\bf R},\alpha,\beta} {\bf S}_{\bf R} \cdot c_{{\bf R} \alpha}^\dag {\boldsymbol \sigma}_{\alpha \beta} c^\pd_{{\bf R} \beta}
\label{ham-realspace}
\end{align}
where $c_{{\bf r},\alpha}^\dag$ creates an electron at lattice site ${\bf r}$ with spin $\alpha$, and ${\bm \sigma}$ is the vector of spin Pauli matrices. As the experimental realizations of such MSH structures have included superconducting Pb(110) \,\cite{nadj-perge-14s602,ruby-15prl197204,pawlak-15npjqi16035} as well as Re(0001) \, \cite{kim-18sa5251,palacio-morales2019} surfaces, which possess a square or triangular surface lattice structure, respectively, we consider below both types of lattices, with $-t$ being the hopping amplitude between nearest-neighbor sites and $\mu$ being the chemical potential.

\begin{figure}[h]
\centering
\includegraphics[width=8.4cm]{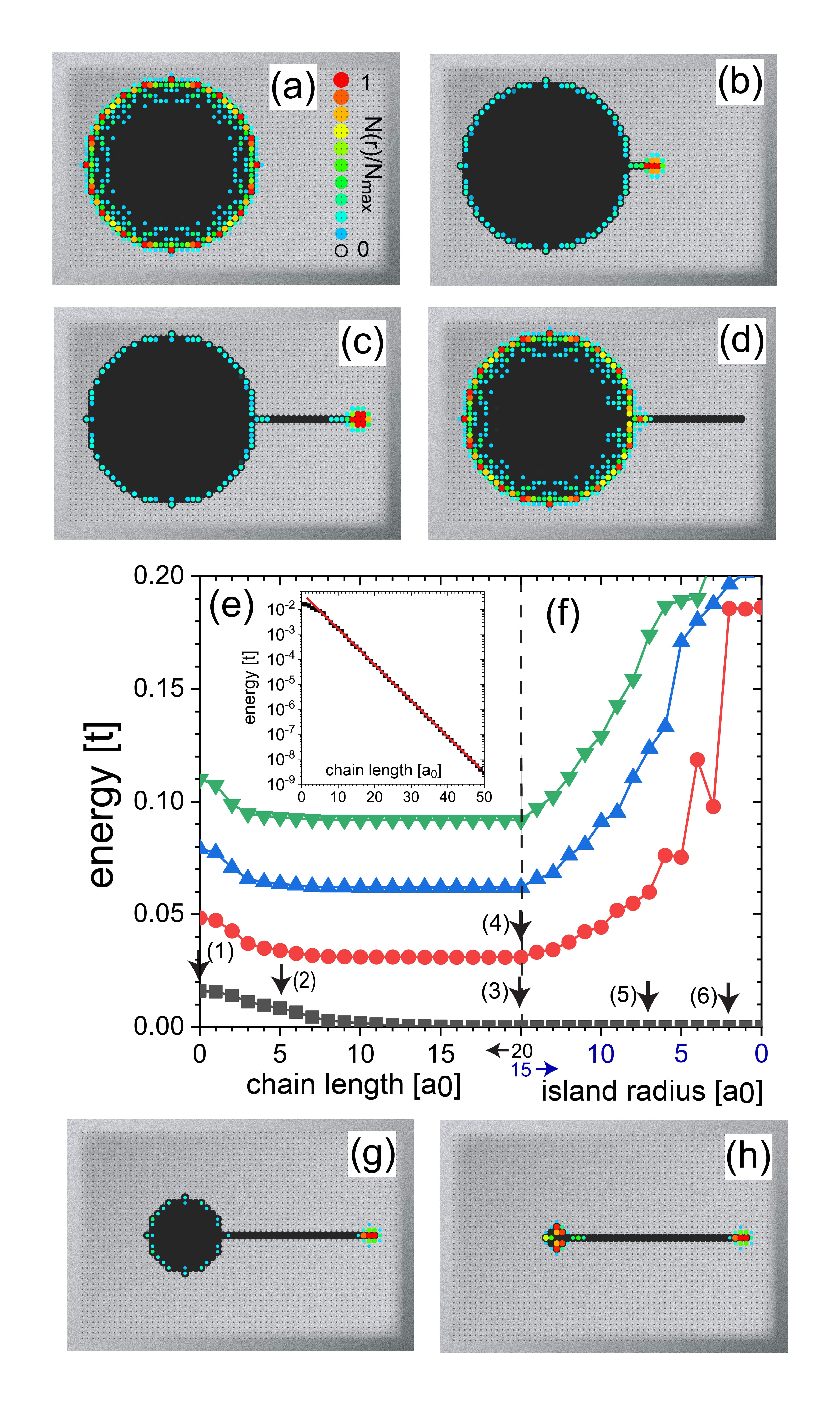}\\
\caption{ MSH hybrid structure consisting of a Shiba island of radius $R=15 a_0$ and chain with parameters $(\mu,\Delta,\alpha,J)=(-4t, 1.2t, 0.45t, 2.6t)$, yielding a topological phase with $C=-1$ for the island (black circles denote sites with magnetic adatoms; $a_0$ is the lattice constant). (a)-(c) LDOS of the lowest energy Majorana mode with increasing chain length, corresponding to arrows (1)-(3) in (e). (d)  LDOS of the second lowest energy Majorana mode, corresponding to arrow (4) in (e). (e) Evolution of the 4 lowest energy levels with increasing chain length. Inset: log-plot of the lowest energy level with increasing chain length, red line represents a linear fit.  (f) Evolution of the 4 lowest energy levels with decreasing island radius. (g),(h) LDOS of the lowest energy Majorana mode with decreasing island radius, corresponding to arrows (5),(6) in (f). The distance between the end of the chain and the center of the island is kept constant. }
\label{fig:fig1}
\end{figure}
Moreover, $\alpha$ denotes the Rashba spin-orbit coupling arising from the breaking of the inversion symmetry at the surface\,\cite{nadj-perge-14s602} with $\boldsymbol \delta$ being the vector connecting nearest neighbor sites. Due to the full superconducting gap, which suppresses Kondo screening, we consider the magnetic moments to be static in nature, such that ${\bf S}_{\bf R}$ is a classical vector representing the direction of the adatom's spin located at ${\bf R}$, and $J$ is its exchange coupling with the conduction electron spin. Unless otherwise stated, we assume an out-of-plane ferromagnetic alignment of the adatoms' magnetic moments.  Finally, we find that the results below are robust over a large range of parameters, as long as the topological phases are not destroyed.

While the topological phases of macroscopic, translationally-invariant systems are well characterized by the topological invariant -- the Chern number -- the topological phases of inhomogeneous \cite{mascot-19arXiv1905.05923} or finite-size systems can be characterized by the spatial Chern number density given by
\begin{equation}
C(\textbf{r}) = \frac{N^2}{2\pi i} {\rm Tr}_{\tau,\sigma} [P[\delta_x P, \delta_y P]]_{\textbf{r},\textbf{r}}\ .
\end{equation}
with the Chern number in real space then given by \,\cite{prodan11jpa113001,prodan17} $C = 1/N^2 \sum_{\textbf{r}}  C(\textbf{r})$. Here, ${\rm Tr}_{\tau,\sigma}$ denotes the
partial trace over spin $\sigma$ and Nambu space $\tau$,
\begin{equation}
\delta_i P= \sum_{m=-Q}^Q c_m e^{-2\pi i m \hat x_i/ N} P e^{2\pi i m \hat x_i / N}\ ,
\end{equation}
 and the projector $P$ onto the occupied bands is given by $P=\sum_{\alpha={\rm occ.}} \ket{\psi_\alpha}\bra{\psi_\alpha}$ for a real-space $N\times N$ lattice. The  $c_m$'s are central finite difference coefficients for approximating the partial derivatives. The coefficients for positive $m$ can be calculated by solving the following linear set of equations for $c = (c_1,\ldots,c_Q)$: $\hat A c = b, A_{ij} = 2j^{2i-1}, b_i = \delta_{i,1}, i,j \in \{1,...,Q\}$.
For negative $m$, we have $c_{-m} = -c_m$. We have taken the largest possible value of $Q= N/2$.  $C(\textbf{r})$ as defined above thus represents the real-space analog of the Berry curvature $\mathcal{F}({\bf k})$, and was previously introduced to discuss the topological phases of Chern insulators\,\cite{bianco-11prb241106}.

\begin{figure*}[t]
\centering
\includegraphics[width=18cm]{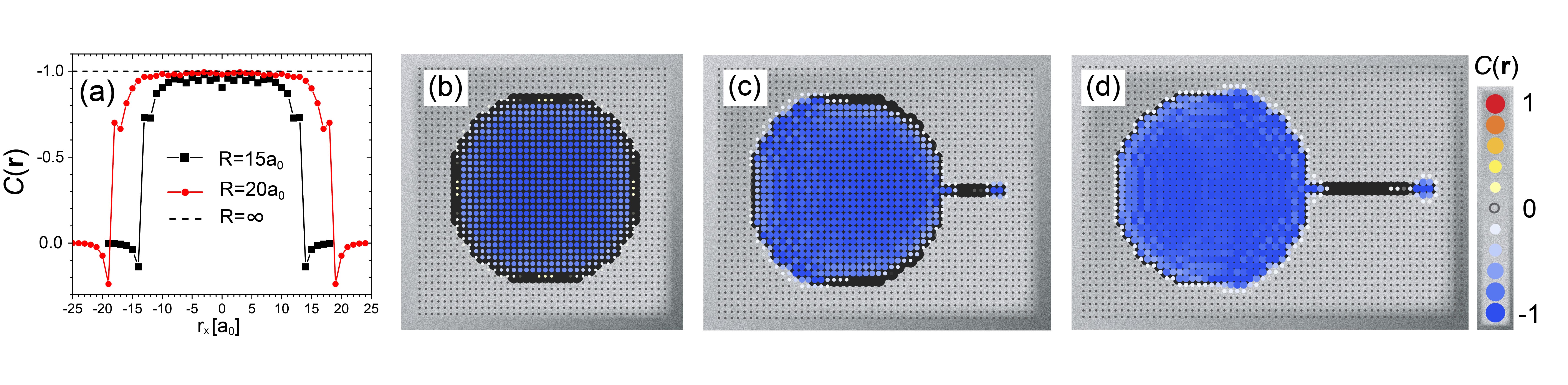}
\caption{(a) Linecut of the Chern density $C({\bf r})$ through Shiba islands with different radii $R$. Spatial plot of $C({\bf r})$ for the MSH structures in Fig.\,\ref{fig:fig1} with chain length (b) $L=0$, (c) $L=10 a_0$ and (d) $L=20 a_0$. Parameters are the same as in Fig.\,\ref{fig:fig1} with $C=-1$.}
\label{fig:Cr}
\end{figure*}

%
%
\section{Dimensional tuning and counting of Majorana modes}

The observation of zero-dimensional (0D) Majorana modes at the ends of Shiba chains, and chiral 1D
Majorana modes at the edges of Shiba islands have raised the question of whether atomic manipulation techniques can be used to tune Majorana modes between these two limits. To address this question, we consider a system in which both the 1D chains and 2D islands of magnetic adatoms induce topological superconductivity, with Chern number $C=-1$ in the latter case. To interpolate between Shiba chains and islands, we attach an increasingly longer chain of magnetic adatoms to a Shiba island, as shown in Figs.~\ref{fig:fig1}(a)-(c). With no chain present, the Shiba island possesses a chiral Majorana mode that is localized at the edge of the island, and forms a dispersing, 1D edge mode that traverses the superconducting gap. Each chiral edge mode is comprised of two Majorana fermions, which for the lowest energy mode are located at small, but finite energy $E=\pm \epsilon$ [Fig.~\ref{fig:fig1}(e)] (this non-zero energy, and the discreteness of the modes arises from the finite size of the island\,\cite{RachelMorr17}). The LDOS of the lowest energy mode is shown in Fig.~\ref{fig:fig1}(a). When a chain is attached to the island, and its length is increased,  spectral weight is transferred from the island's lowest energy Majorana edge mode to the end of the chain [Fig.~\ref{fig:fig1}(b)]. When the chain is sufficiently long [Fig.~\ref{fig:fig1}(c)] it possesses a localized Majorana fermion at its end, while a second Majorana fermion remains delocalized along the edge of the island. A similar spatial structure of the zero-energy LDOS was also found in plaquette--nanowire hybrid system \cite{Kob18}. Note that the spatially integrated spectral weight of the zero-energy state is exactly split between the dispersive Majorana edge fermion and the bound state at the end of the chain. Concomitant with the increasing separation between these two Majorana fermions, the energy of the lowest energy states decreases smoothly and monotonically [Fig.~\ref{fig:fig1}(e)], implying that the system remains in a topological phase throughout this evolution, i.e., without undergoing a phase transition. Moreover, as expected for two Majorana fermions localized at the end of a chain, the log-plot of their energy shown in the inset of Fig.~\ref{fig:fig1}(e) reveals that it decreases exponentially with increasing length of the chain.  At the same time, the higher energy Majorana modes remain entirely localized along the edge of the island, as shown in Fig.~\ref{fig:fig1}(d) for the second lowest energy state. We note that as there is no zero-energy crossing of the lowest energy states with increasing chain length, no parity transitions occurs \,\cite{ben-shach-15prb045403}. To finally arrive at a purely one-dimensional chain, we next reduce the radius of the Shiba island, while keeping the distance between the end of the chain and the center of the island fixed. In Fig.~\ref{fig:fig1}(f), we present the evolution of the four lowest energy states with decreasing island radius, $R$ [the purely 1D chain corresponds to the right hand side of Fig.~\ref{fig:fig1}(f)]. As expected, we find that the lowest energy state, corresponding to the Majorana fermions located at the end of the chain and around the edge of the island, remains located near zero energy with decreasing $R$, while the other states move up in energy. The corresponding lowest energy LDOS for two values of $R$ are shown in Figs.\ref{fig:fig1}(g),(h), demonstrating that the Majorana fermions remain located at the ends of the chain with decreasing $R$. The smooth evolution of the lowest energy states shown in Fig.~\ref{fig:fig1}(f) demonstrates that the system can be adiabatically tuned, i.e., without undergoing a phase transition, between the island-chain structure shown in Fig.~\ref{fig:fig1}(c), and a 1D Shiba chain. This implies that the entire transition from the 2D Shiba island in Fig.~\ref{fig:fig1}(a) [corresponding to the left hand side in Fig.~\ref{fig:fig1}(e)] to the 1D chain [corresponding to the right hand side of Fig.~\ref{fig:fig1}(f)] is adiabatic, without an intermittent topological phase transition.

Further evidence for the adiabatic evolution of the system comes from considering the Chern number density, $C({\bf r})$, of the MSH structure. $C({\bf r})$ can be employed to
characterize the topological nature of finite-size (or disordered \cite{mascot-19arXiv1905.05923}) systems, as it adiabatically connects to the quantized Chern number of translationally invariant, macroscopic systems. To demonstrate this, we compare in Fig.~\ref{fig:Cr}(a) the Chern number density along a line cut through Shiba islands of different radii, with $C({\bf r})$ of an infinitely large system. Already for an island with radius $R=15a_0$, $C({\bf r})$ in the center of the island is close to that of the infinitely large system, an agreement which improves with increasing island radius. At the same time, the existence of chiral edge modes in these island [as shown in Fig.~\ref{fig:fig1}(a)] provide further, independent evidence for their topological nature. Consistent with this, we find that a vanishing $C({\bf r})$ inside the island coincides with the absence of any chiral edge modes, thus reflecting a topological trivial phase.  We can thus conclude that the Chern number density is a suitable quantity to characterize the topological nature of finite-size Shiba islands.

Having established this, we can now employ $C({\bf r})$ to further demonstrate that the MSH system shown in Figs.~\ref{fig:fig1}(a)-(c) can be adiabatically tuned between 1D and 2D without undergoing a phase transition. To this end, we present in Figs.~\ref{fig:Cr}(b)-(d) the Chern number density for the MSH structure with different chain lengths, ranging from an island without chain [Fig.\,\ref{fig:Cr}(b)], to islands with chain lengths $L=10$ [Fig.\,\ref{fig:Cr}(c)] and $L=20$ [Fig.\,\ref{fig:Cr}(d)]. Throughout this evolution, $C({\bf r})$ not only remains close to $-1$ inside the Shiba island, but the end of the chains also exhibits some non-zero negative value for $C(\textbf{r})$, again demonstrating the adiabatic evolution of the MSH structure. Thus, the results shown in Figs.~\ref{fig:fig1} and \ref{fig:Cr} demonstrate that it is not only possible to adiabatically tune between 1D and 2D topological superconductivity via atomic manipulation (and hence spatially separate Majorana fermions without the creation of magnetic vortices\,\cite{alicea-12rpp076501}), but also to design a single system exhibiting both localized and dispersive Majorana zero modes.

The above results open a new path to the long sought goal for identifying the topological invariant --- the Chern number --- through measurements, as attaching chains to a magnetic island via atomic manipulation allows one to count the Chern number.
\begin{figure}[t!]
\centering
\includegraphics[width=9cm]{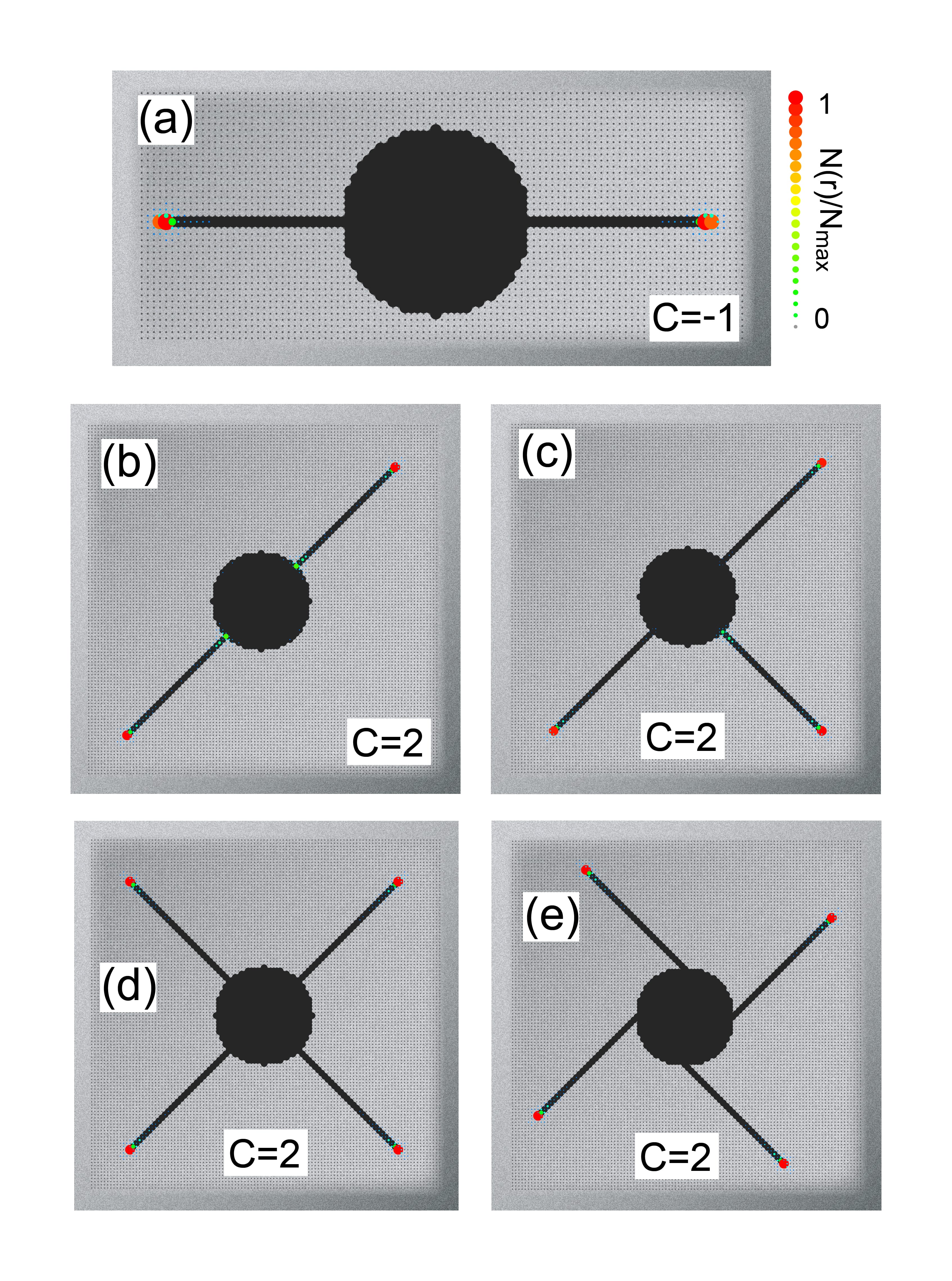}
\caption{(a) LDOS of the lowest energy Majorana mode for a Shiba island with $C=-1$ and two chains of length $L=30 a_0$ attached (same parameters as in Fig.~\ref{fig:fig1}). (b)-(e) MSH structure with $C=2$ and parameters $(\mu,\Delta,\alpha,J)=(-0.5t, 0.7t, 0.45t, 2t)$. Lowest energy LDOS for a Shiba island with radius $R=15a_0$ and with (b) two, (c) three, and (d) four chains of length $L= 31 a_0$ attached. (e) The points where the chains are attached are rotated by $45 ^\circ$ from (d).}
\label{fig:fig2}
\end{figure}
To demonstrate this, we attach a second chain to the island--chain hybrid system of Fig.~\ref{fig:fig1}(c), and present the resulting lowest energy LDOS in Fig.~\ref{fig:fig2}(a). As expected, we find that by attaching a second chain, the dispersive Majorana mode moves from the edge of the Shiba island [see Fig.~\ref{fig:fig1}(c)] to the end of the second chain where it forms a bound state. Thus, attaching two chains to a Shiba island that is in a topological phase with $C=\pm 1$ relocates the zero-energy Majorana modes from the edge of the island to the end of the chains, as shown in Fig.~\ref{fig:fig2}(a). Coincidentally, this result also demonstrates that the localization of Majorana modes at the end of a chain is insensitive to the particular shape of the chain in its middle. Next, we consider a Shiba island that is in a topological superconducting phase with Chern number $C=2$, implying that the island possesses two degenerate chiral Majorana edge modes. We note that these two modes are topologically protected from combining into a complex Dirac fermion \cite{RachelMorr17}, due to the absence of any coupling between them. When two chains are attached to such an island with $C=2$, one of the Majorana modes is separated into two zero energy Majorana fermions which are located at each end of the two chains, as shown in Fig.~\ref{fig:fig2}(b) [we note that for this particular set of parameters, the chains are only in a topological phase when they are oriented along the diagonal, but not when they are oriented along the bond directions, which is opposite to the case considered in Fig.~\ref{fig:fig1}]. At the same time, the second Majorana mode remains localized along the edge of the Shiba island, with large spectral weight concentrated near those points along the edge where the chains are attached. This result is qualitatively different from the $C=-1$ case where in the presence of two chains, no low-energy Majorana mode remains along the edge of the Shiba island [see Fig.~\ref{fig:fig2}(a)]. When a third chain is attached to the island [see Fig.~\ref{fig:fig2}(c)], the second Majorana mode is spatially split into two Majorana fermions, one that is located at the end of the third chain, and one that remains located along the edge of the island. Only when four chains are attached to the island [Fig.~\ref{fig:fig2}(d)] we find that four zero-energy Majorana fermions (arising from the two lowest energy Majorana modes of the island) are located at the end of the chains, with no zero-energy mode remaining along the edge of the island. We note that this result does not depend on the particular locations where the chains are attached to the island, as shown in Fig.\,\ref{fig:fig2}(e), being a variant of the geometry in Fig.\,\ref{fig:fig2}(d).

The results shown in Fig.~\ref{fig:fig1}(c) and Fig.~\ref{fig:fig2} suggest a new real space approach to detecting the Chern number of a two-dimensional topological superconductor through atomic manipulation: if spectral weight for a zero-energy state remains located at the edge of the island when $N-1$ chains are attached, but vanishes for $N$ chains, then the Chern number of the 2D topological superconductors is given by $|C|=N/2$. Our results also imply that by attaching chains to the island, one can effectively change the number of Majorana zero modes in the island, and in particular tune the island to an even or odd number of zero-energy modes.

\begin{figure}[t]
\centering
\includegraphics[width=9cm]{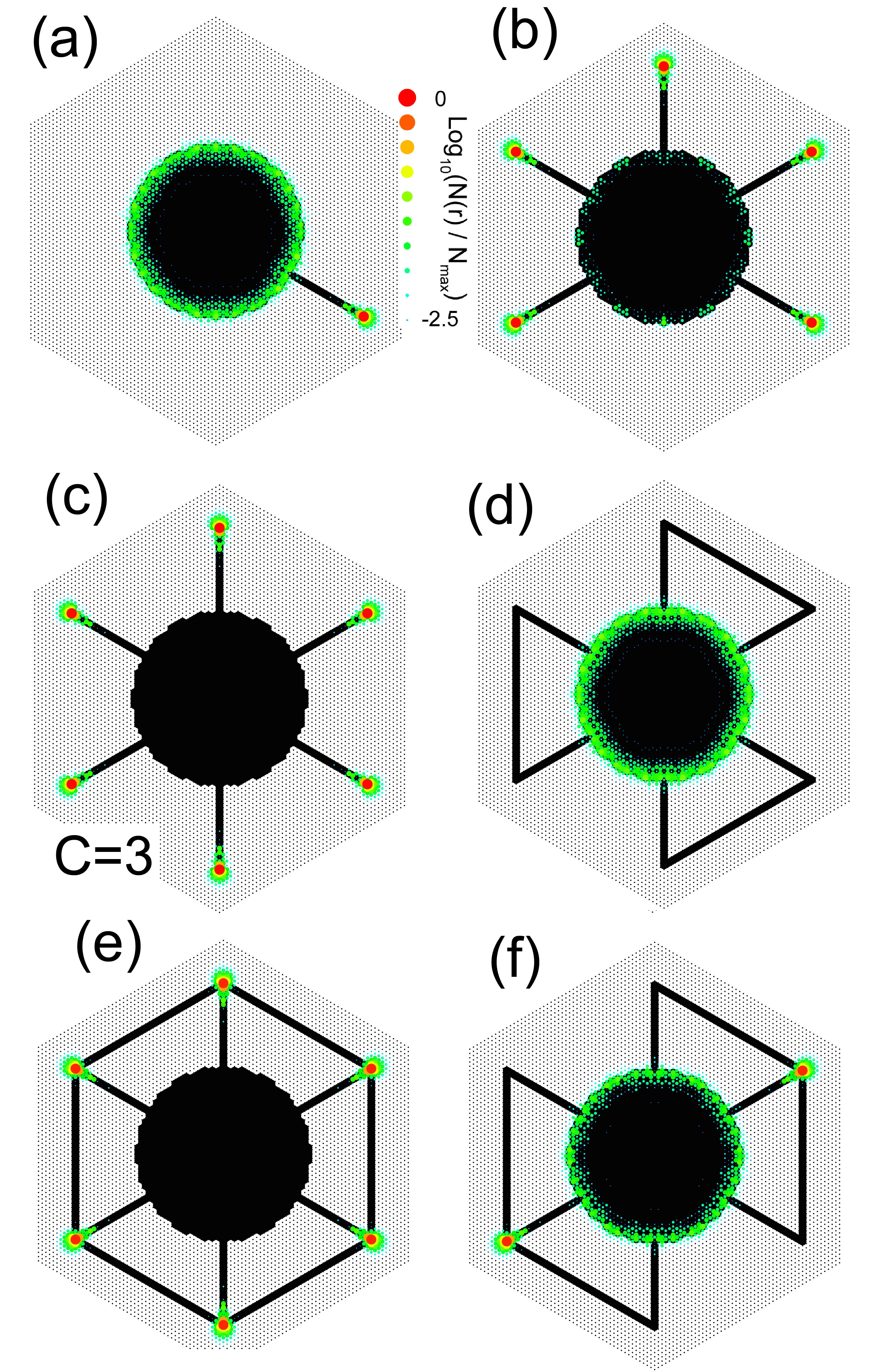}
\caption{Shiba island with radius $R=20a_0$ on a triangular lattice with parameters $(\mu,\Delta,\alpha,J)=(0.4, 1.2, 0.45, 2.6 )t$ yielding a topological phase with $C=3$. Zero energy LDOS for the island with (a) one, (b) five and (c) six chains of length $L=20a_0$ attached. Zero energy LDOS for the island with (d) the ends of two neighboring chains connected, giving rise to a windmill-like structure and junctions with an even number of chains (even junction), (e) the ends of neighboring chains connected, giving rise to a spiderweb-like structure and junctions with an odd number of chains (odd junction), and (f) the ends of two opposite chains connected with the ends of their two neighboring chains giving rise to a TIE fighter-like structure with both even and odd junctions.}
\label{fig:fig3}
\end{figure}
To demonstrate that these findings also hold for higher Chern numbers and different lattice structures, we next consider a Shiba island in the $C=3$ phase located on a triangular lattice, and present in Figs.~\ref{fig:fig3}(a)-(c) the resulting lowest-energy LDOS for an increasing number of chains. Such an island possesses 3 degenerate Majorana modes, or 6 Majorana fermions. As expected, we find that with each added chain, a Majorana fermion is removed from the Shiba island and relocated to the end of the chain, leading to a decrease in the spectral weight around the edge of the island, as shown in Figs.~\ref{fig:fig3}(a),(b). When six chains are attached to the island, all low-energy Majorana modes are removed from the edge of the island [Fig.~\ref{fig:fig3}(c)]. These results again fully support the validity of the real space approach to counting the Chern number introduced above, as for $N=6$ chains no zero-energy spectral weight remains around the edge of the island, implying that it is in the $C=N/2=3$ phase.

Atomic manipulation can be employed to further tune the location of the Majorana fermions between the end of the chains and the edge of the island. For example, connecting the ends of two neighboring chains by another chain -- giving rise to the windmill-like structure shown in Fig.~\ref{fig:fig3}(d) -- leads to junctions where an even number of chains meet (even junctions). Such junctions cannot sustain the existence of Majorana fermions \cite{Bjo16}, such that all Majorana fermions are relocated back to the edge of the Shiba island. On the other hand, connecting the ends of all chains with the ends of their nearest neighbor chains [see Fig.~\ref{fig:fig3}(e)] gives rise to six junctions where an odd number of chains meet (odd junctions). Such junctions can sustain the existence of a Majorana fermion, such that all Majorana fermions remain located at the end of the chains, i.e., at the junctions. Finally, connecting the ends of two opposite chains with the ends of their two neighboring chains  -- leading to the TIE fighter-like structure shown in Fig.~\ref{fig:fig3}(f) -- produces 4 even and 2 odd junctions, such that 4 Majorana fermions are relocated to the edge of the island, and 2 Majorana fermions remain located at the odd junctions.

\begin{figure}[t!]
\centering
\includegraphics[width=9cm]{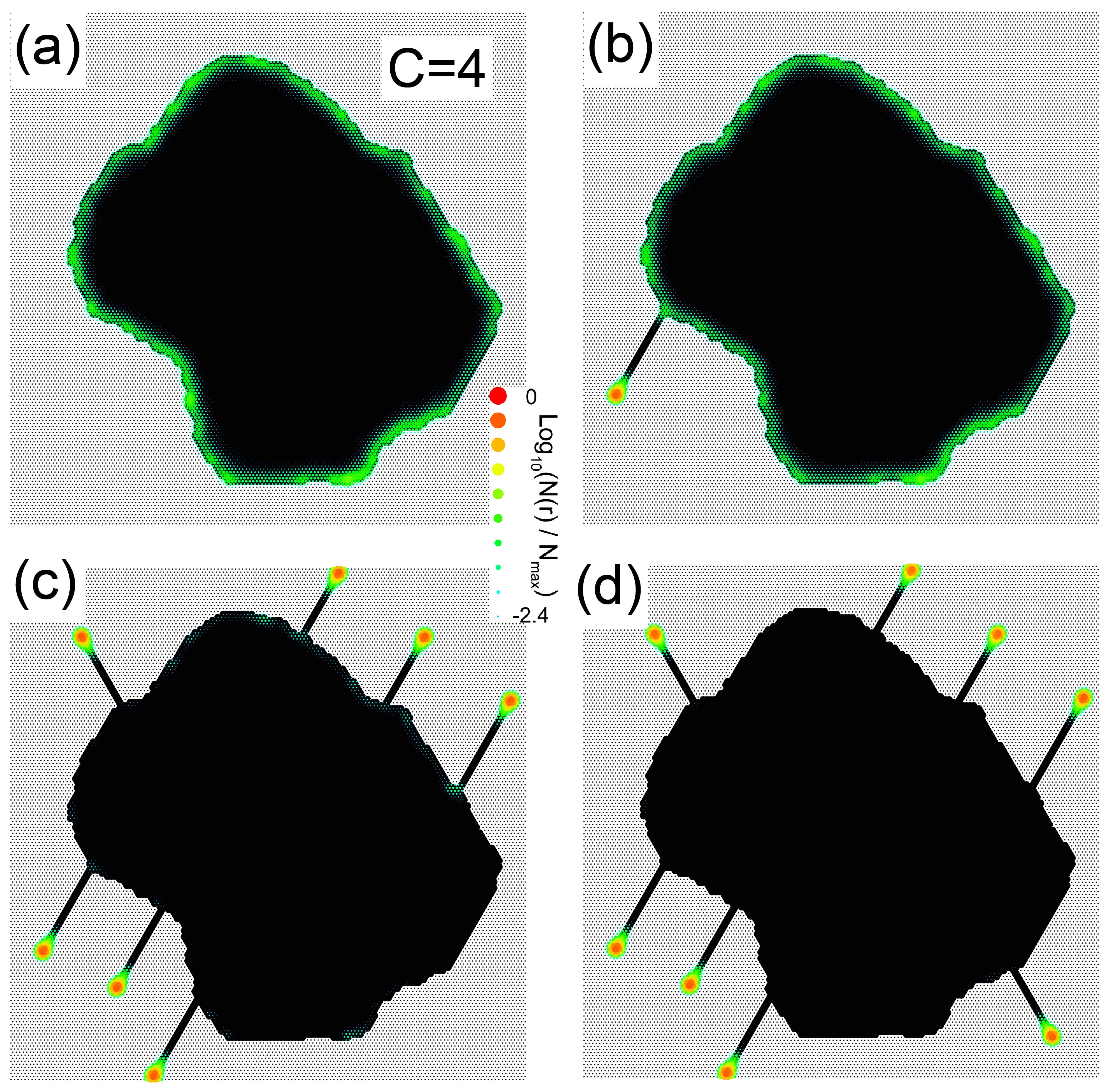}
\caption{Experimentally realized Fe/Re-O$(2 \times 1)$ MSH structure (see Ref.~\onlinecite{palacio-morales2019}) with $C=4$. LDOS of zero-energy Majorana modes with (a) no attached chain, (b) one, (c) seven, and (d) eight attached chains. Hamiltonian and parameters for this 10-band model are given in the supplemental material of Ref.~\onlinecite{palacio-morales2019} and appendix \ref{sec:para}.}
\label{fig:fig4}
\end{figure}
To demonstrate that dimensional tuning and real space counting of Chern numbers can also be achieved in experimentally relevant MSH structures, we consider an MSH structure, described by a 10-band model, that was recently employed to explain the emergence of topological superconductivity in Fe islands deposited on a Re(0001)-O(2x1) surface\,\cite{palacio-morales2019} (for the Hamiltonian describing this system, see the supplemental material of Ref.~\onlinecite{palacio-morales2019} and appendix Sec.\ref{sec:para}). The Fe island considered below possesses the same form and spatial extent as the experimentally studied one \,\cite{palacio-morales2019}. However, while the experimentally observed island is in a topological phase with $C=20$, we have slightly altered the parameters to tune the island into a topological phase with a smaller Chern number of $C=4$. In this case, the above counting argument predicts that only $N=8$ chains need to be attached to the island to remove all zero-energy Majorana fermions from the edge of the island, rather than the $N=40$ chains required in the $C=20$ phase.

As the island is in the $C=4$ topological phase, it exhibits four chiral Majorana modes along the edge of the island, with the corresponding $E=0$ LDOS shown in Fig.\,\ref{fig:fig4}\,(a). As before we find that a single chain of magnetic adatoms attached to the island possesses a Majorana bound state that is localized at the end of the chain, with all other Majorana fermions remaining located along the edge of the island, as follows from the zero-energy LDOS shown in Fig.~\ref{fig:fig4}(b). For an island with 7 chains attached [see Fig.~\ref{fig:fig4}(c)], substantial spectral weight has been removed from the edge of the island as 7 Majorana bound states are now located at the end of the chains, with only a single chiral Majorana fermion remaining located along the edge of the island. We note that as expected the integrated spectral weight along the edge of the island is the same as that around each end of the chains. However, as the spectral weight of the Majorana fermion is extended over the entire length of the edge, the peak intensity in the LDOS along the edge is significantly smaller than that at the end of the chains [see Fig.~\ref{fig:fig4}(c)]. Finally, adding an 8th chain to the island relocates the last Majorana fermion to the end of the chain, leading to a vanishing zero-energy LDOS at the edge of the island [Fig.\,\ref{fig:fig4}\,(d)]. We therefore conclude that proposed real space approach to counting the Chern number holds even for experimentally relevant MSH structures, thus further supporting its validity.

The above results also suggest how the dimensional tuning of Majorana fermions between 1D and 2D can be generalized to MSH structures with Chern numbers different from $C=\pm 1$ which was discussed in Fig.~\ref{fig:fig1}. Specifically, we find that it is possible to use atomic manipulation techniques to adiabatically tune between a Shiba island with $|C|$ chiral Majorana modes, and a network of chains that host $2|C|$ localized Majorana bound states, as shown in Figs.~\ref{fig:fig2}(a), \ref{fig:fig2}(d), \ref{fig:fig3}(c),(e) and \ref{fig:fig4}(d). Atomic manipulation techniques thus provide a promising new approach to the quantum engineering of Majorana fermions.

%
%
\section{Conclusions}
In conclusion, we have demonstrated that it is possible to use atomic manipulation techniques to adiabatically tune MSH structures between 1D and 2D topological superconducting phases. Specifically, we showed that while two-dimensional chiral topological superconductors (with $Z$ classification) and one-dimensional topological superconductors (with $Z_2$ classification) are in different universality classes, the system does not undergo a phase transition if one transforms a 2D Shiba island via a hybrid chain-island structure into a 1D chain by adding or removing adatoms. Moreover, by attaching Shiba chain networks to Shiba islands, we showed that one can arbitrarily transform chiral Majorana edge modes into localized Majorana bound states, and vice versa. This, in turn, opens a new real space approach to counting the Chern number of topological superconductors. In particular, when a Shiba island is in a topological phase with Chern number $C$, then the spectral weight of the zero-energy chiral Majorana edge modes completely vanishes when $2|C|$ chains are attached to it, as the Majorana modes are transformed into Majorana bound states localized at the end of the chains. We have explicitly demonstrated these results for a series of Chern numbers ($C=1, 2, 3, 4)$ and for different lattice geometries, but also in an experimentally relevant MSH structure that was recently successfully employed to explain the emergence of chiral Majorana edge modes in Fe/Re-O$(2 \times 1)$ MSH system \cite{palacio-morales2019}. Finally, we demonstrated that the topological nature of MSH structures can be characterized through the Chern number density, which adiabatically connects between finite-size MSH structures and translationally invariant, macroscopic systems that are characterized by an integer Chern number. We note in closing that the evolution of topological superconductivity between 1D and 2D was previously studied by considering a $p_x+ip_y$ superconductor whose geometry was tuned from a rectangle to a narrow strip, quasi one-dimensional system of length $L$ and varying width $W$ \cite{Pot10}. We find, however, that such a system undergoes a topological phase transition when $W$ is varied, such that the 1D and 2D limits cannot be adiabatically connected.

\section{Acknowledgements}
The authors would like to thank H.\ Kim, A. Kubetzka, T. Posske, K. von Bergmann, M.\ Voj\-ta and R.\ Wiesendanger for stimulating discussions.
This work was supported by the U. S. Department of Energy, Office of Science, Basic Energy Sciences, under Award No. DE-FG02-05ER46225 (E.M., S.C., and D.K.M.) and through an ARC Future Fellowship (FT180100211) (S.R.).

\appendix

\section{Parameters for the Fe/Re-O$(2 \times 1)$ MSH structure}
\label{sec:para}

To reduce the Chern number from experimentally relevant case of $C=20$ for experimentally realized Fe/Re-O$(2 \times 1)$ structure in Ref.\cite{palacio-morales2019} to $C=4$ discussed in Fig.~\ref{fig:fig4}, the following parameters were changed from those given in the supplemental material of Ref.\,\onlinecite{palacio-morales2019}: $\Delta_{Re} = 33$ meV, $\mu_{Fe}=-10.96$ meV,  $\lambda_{Fe} = 0.64$ meV, $\alpha_{Fe}  = 3.36$ meV, and $\Delta_{Fe}  = 3.28$ meV.

\bibliographystyle{Science}

\end{document}